\documentclass[trackchanges,twocolumn,astrosymb]{aastex701}

\definecolor{darkperiwinkle}{RGB}{102, 102, 128}

\begin{document}

\title{A Black Hole is Born: 3D GRMHD Simulation of Black Hole Formation from Core-Collapse}

\author[0000-0002-7232-101X]{Goni Halevi}
\affiliation{Department of Physics, Illinois Institute of Technology, 10 West 35th Street, Chicago, IL 60616, USA}
\affiliation{Center for Interdisciplinary Exploration and Research in Astrophysics (CIERA), Northwestern University, Evanston, IL 60201, USA}
\email[show]{ghalevi@illinoistech.edu}

\author[0000-0002-5109-0929]{Swapnil Shankar}
\affiliation{Faculty of Mathematics, Informatics and Natural Sciences, University of Hamburg, Gojenbergsweg 112, 21029 Hamburg, Germany}
\affiliation{Department of Physics and Astronomy, University of Tennessee Knoxville, Knoxville, TN 37996, USA}
\affiliation{GRAPPA, Anton Pannekoek Institute for Astronomy, Institute of High-Energy Physics, and Institute of Theoretical Physics, University of Amsterdam, Amsterdam, The Netherlands}
\email{swapnilshankar1729@gmail.com}

\author[0000-0002-9371-1447]{Philipp M\"{o}sta}
\affiliation{GRAPPA, Anton Pannekoek Institute for Astronomy, Institute of High-Energy Physics, and Institute of Theoretical Physics, University of Amsterdam, Amsterdam, The Netherlands}
\email{p.moesta@uva.nl}

\author[0000-0003-1424-6178]{Roland Haas}
\affiliation{Department of Physics and Astronomy, University of British Columbia, 6224 Agricultural Road, Vancouver, British Columbia, Canada }
\affiliation{National Center for Supercomputing Applications, University of Illinois, 1205 W Clark St, Urbana, Illinois, USA}
\affiliation{Department of Physics, University of Illinois, 1110 West Green St, Urbana, Illinois, USA}
\email{rhaas@phas.ubc.ca}

\author[0000-0002-4518-9017]{Erik Schnetter}
\affiliation{Perimeter Institute for Theoretical Physics, Waterloo, Ontario, Canada}
\affiliation{Department of Physics and Astronomy, University of Waterloo, Waterloo, Ontario, Canada}
\affiliation{Center for Computation \& Technology, Louisiana State University, Baton Rouge, Louisiana, USA}
\email{eschnetter@perimeterinstitute.ca}

\begin{abstract}

We present the first three-dimensional, fully general-relativistic magnetohydrodynamic (3D GRMHD) simulation of a black hole (BH) formed from the collapsed core of a massive star. The ability to self-consistently capture the birth of a compact remnant in 3D is crucial for modeling natal BH properties (including masses, spins, and kicks), which are of particular interest in the era of gravitational wave astronomy. However, such simulations have remained elusive due to extreme computational challenges and demands. We employ the GPU-accelerated dynamical-spacetime GRMHD code \texttt{GRaM-X} to follow the collapse, core-bounce, shock propagation, and eventual BH formation of a massive stellar progenitor in full 3D. We initialize our simulation by mapping a one-dimensional (1D) model of a star with a zero-age-main-sequence mass of $45M_\odot$ to 3D. We use the core rotation velocity expected from stellar evolution modeling and a relatively weak dipolar magnetic field. The collapsing core drives a shock that reaches a maximum radius of roughly 170 km before stalling and does not lead to a successful explosion. The proto-neutron star accretes matter before collapsing to form a BH $t_\mathrm{BH} \approx 325~\mathrm{ms}$ after core-bounce. The time of BH formation and initial BH mass are remarkably similar to those obtained with \texttt{GR1D}, a 1D general-relativistic neutrino-hydrodynamics code, to which we compare our results. We track the horizon of the newborn BH after formation and calculate a steady kick velocity of $\left<v_\mathrm{kick}\right> \approx 72~\mathrm{km/s}$ and a mass of $M_\mathrm{BH} \approx 2.62 M_\odot$, which is still rising at the end of the simulation.

\end{abstract}


\section{Introduction} \label{sec:intro}

It is regarded as common knowledge that the collapsed cores of massive (zero-age-main-sequence mass $M_\mathrm{ZAMS} \gtrsim 8M_\odot$) stars give birth to compact objects. The general picture is one in which, depending on the mass of the core at the onset of runaway gravitational collapse, a star will leave behind either a neutron star (NS) or a stellar-mass black hole (BH). A proto-NS (PNS) is born alongside a core-collapse supernova (CCSN) explosion, while a newborn BH may result from either a successful or failed CCSN. However, direct observational links between the core-collapse process and these compact objects have long been elusive. CCSNe are shrouded in optically thick ejecta and dust, which obscures any central remnant long after the explosion. Meanwhile, the search for failed SNe is active, but the signature of a disappearing star and the BH it may leave behind remains challenging to identify conclusively. 

To date, there are only a handful of confirmed observational connections between compact remnants and the cataclysmic events that produce them. Older supernova remnants like the Crab Nebula provide an opportunity to detect the compact object left behind by the explosion after the ejecta has expanded to become optically thin. The long-known existence of the Crab Pulsar (PSR B0531+21) confirms that the historical CCSN associated with this remnant, SN 1054, created an NS. More recently, there have been claims that SN 2022jli, a stripped-envelope (Type Ic) SN with unusual periodic oscillations in its light curve, is evidence of a binary system including a compact bound remnant produced by the explosion \citep{SN2022jli_1,SN2022jli_2}. A cooling NS has also been proposed as responsible for narrow infrared emission lines observed with JWST in the nearby, well-studied SN 1987a \citep{1987a_JWST}. Direct associations such as these collectively provide only a small window into the statistical nature of the core-collapse to compact object connection.

Building a complete, population-level understanding of compact objects at birth, including the statistical distributions of natal masses, spins, and kicks, is particularly lucrative in the era of gravitational wave (GW) astronomy. Accurate predictions for and interpretations of compact binary mergers like those now detected in large numbers by the LIGO-Virgo-KAGRA \citep[LVK;][]{LVKLRR,LVKpop} collaboration require knowledge of the underlying population. Modeling how core-collapse in massive stars maps to natal properties of BHs, including whether they are likely to become unbound in a binary, is of great interest to the LVK and binary population synthesis communities. CCSNe themselves are expected to produce ``burst'' GW signatures that, if the SN occurs sufficiently nearby, will be detectable with current and future GW observatories \citep[e.g.,][]{SNGW1,SNGW2,SNGW3,SNGW4,SNGW5}.

Numerical models of compact object formation have required a trade-off: we either use simulations with a simplified treatment of general relativity, which cannot capture BH formation with full accuracy, or spherically symmetric ones that can form BHs and explore a larger parameter space, at the steep cost of not including multidimensional effects. The complex physics of CCSN explosions-- including microphysical equations of state (EoS), nuclear reactions, neutrino transport, and (magneto-)hydrodynamics (MHD)-- and their inherent asymmetries render full-physics simulations challenging and expensive. Full 3D simulations that include all these ingredients sacrifice spacetime evolution, making it impossible to capture BH formation. The compromise is to extrapolate to the final remnant without actually simulating its creation. For example, the multigroup, multidimensional radiation hydrodynamics code \texttt{FORNAX} \citep{fornax} is one of a few that successfully models CCSNe in 3D \citep{fornax3d} but employs a monopole approximation to general relativity (GR). Recent results based on 20 \texttt{FORNAX} simulations have led to a proposed theory for NS and BH birth \citep{burrows_NSBHtheory} that is much more complicated than the classic picture \citep[e.g., that of][]{fryer_COform}. They track the long-term post-BH formation evolution by remapping to another code and find intriguing consequences for the different channels of formation for stellar-mass BHs \citep{burrows_BHchannels}.

In addition to lacking complete GR, some codes \citep[e.g.,][]{fornax,chimera,vertex} lack MHD and are thus limited to exploring non-magnetized models and purely neutrino-driven explosions. However, one of the most interesting cases of self-consistent BH formation from massive stars is that of long-duration gamma-ray burst (lGRB) progenitors. A dominant theory for lGRBs is that they form from collapsars, which require magnetized, rapidly rotating stellar cores. These events are a proposed source of the heaviest elements in the Universe-- those formed through the rapid neutron capture (\textit{r}-process) nucleosynthesis channel. State-of-the-art collapsar GRMHD models \citep[e.g.,][]{gottlieb22a} start from embedding a BH in a rotating, magnetized stellar envelope and follow the accretion, but results are inconclusive about whether realistic stellar conditions would lead to jets that match observed lGRB properties \citep{2023ApJ...944L..38H,issa2025,2025arXiv250202077S}. Although other groups use 3D MHD to simulate magneto-rotational explosions, they are either special-relativistic rather than full GR \citep[e.g.,][]{obergaulinger:17a,Obergaulinger_2021,Bugli_2021,Powell_2023} or otherwise do not run until BH formation \citep[e.g.,][]{mosta14,Mosta:2017geb,halevi:18,Shibagaki_2024}, requiring extrapolation to determine the final remnant properties.

Meanwhile, 1D simulations with dynamical spacetime, such as those performed with \texttt{GR1D} \citep{oconnor10,oconnor15} can capture BH formation and include sophisticated neutrino transport. They also have the benefit of being computationally inexpensive, enabling large parameter studies. However, the assumption of spherical symmetry sacrifices some of the physics that affects natal properties, and until now, the accuracy of their predictions had not been verified \citep[though see][]{2018JPhG...45j4001O}. They are fundamentally unable to probe inherently asymmetrical signatures of BH formation, such as natal kicks and GW signatures, despite recent progress in implementing multidimensional physics \citep{2020ApJ...890..127C,2021ApJ...912...29B}. Recently, \citet{2024MNRAS.533L.107K} bridged these two approaches-- inexpensive 1D numerical relativity and 3D radiation hydrodynamics-- by conducting axisymmetric numerical relativity radiation GRMHD simulations of extremely massive core-collapse. By sacrificing a third spatial dimension, they were able to simulate BH formation from a $M_\mathrm{ZAMS}=70M_\odot$ stellar model while varying rotation and magnetic fields.

In this Letter, we present the first self-consistent simulation of BH formation from massive stellar core-collapse in 3D GRMHD. We describe our progenitor model and numerical tools in \S \ref{sec:methods}. In \S \ref{sec:results}, we describe the results of our simulation and compare to the 1D evolution of the same stellar model. We end by placing our results in context and laying out future directions in \S \ref{sec:discussion}. 

\section{Methods and Initial Conditions} \label{sec:methods}

\subsection{Stellar Progenitor Model} \label{sec:model}
We simulate the 3D evolution of a pre-collapse 1D stellar model with a zero-age-main-sequence (ZAMS) mass of $M_\mathrm{ZAMS} = 45 M_\sun$, as presented by \citet[][hereafter AD20]{aguileradena20}. The model, along with the other stellar models in AD20, was calculated using the open-source 1D stellar evolution Modules for Experiments in Stellar Astrophysics, version 10398 \citep[\texttt{MESA};][]{Paxton2011,Paxton2013,Paxton2015,Paxton2018}. The star was initialized with (1) an equatorial rotational velocity of $600~\mathrm{km}~\mathrm{s}^{-1}$ and (2) a low metallicity of (1/50)$Z_{\sun}$, where $Z_\sun$ is the solar metallicity with abundances scaled from \citet{grevesse96}. This relatively rapid rotation leads to effective mixing and quasi-chemically homogeneous evolution. The fast-rotating pre-collapse core with a hydrogen- and helium-depleted envelope makes this model a potential progenitor of long-duration gamma-ray bursts (lGRBs) \citep[][AD20]{2018ApJ...858..115A}. The \texttt{MESA} model ends at the onset of core-collapse (defined as the time when the core infall velocities first surpass 1000 km s$^{-1}$) at which point we map the result from the stellar evolution code to model the collapse process itself with \texttt{GR1D} and \texttt{GRaM-X}.

We choose this $M_\mathrm{ZAMS} = 45 M_\sun$ model to evolve in 3D GRMHD because, at the onset of core-collapse, it is predicted by some indicators to result in a failed SN and form a BH without driving an explosion. More importantly, it was evolved in 1D by \citet{2023ApJ...944L..38H} and resulted in prompt BH formation (see \S \ref{sec:GR1D} and \ref{sec:comp_1d}). The pre-collapse star has a mass of $M_\mathrm{pre-collapse} = 33.59 M_\odot$. The mass loss throughout this star's evolution is driven by rotation and enhanced by neutrino-driven contraction. We note that the true nature and rate of mass loss in such low-metallicity, rapidly rotating stars are debated \citep[e.g.][]{2015A&A...581A..15S,2020A&A...634A..79S}. 

The most basic, historically applied criterion for explodability is the single-parameter core-compactness test, which is motivated by hydrodynamic simulations of neutrino-driven SNe. The  core-compactness $\xi_M$ is defined as
\begin{equation}
    \xi_M = \frac{M/M_\sun}{R(M)/1000~\mathrm{km}}, \label{eqn:compactness}
\end{equation}
where $R(M)$ is the radius of the enclosed baryonic mass $M$ \citep{2011ApJ...730...70O}. The value of this parameter in a pre-collapse star is commonly used as an indicator of `explodability'-- whether or not a non-rotating stellar core will lead to a successful neutrino-driven explosion. It is typically measured at a mass coordinate of 2.5 $M_\sun$, which generally corresponds to an infall velocity of roughly 1000 km s$^{-1}$. Non-rotating cores with $\xi_{2.5} \lesssim 0.45$ are considered likely to explode, as calibrated by core-collapse simulations  \citep{sukhbold2014}. At the end of its \texttt{MESA} evolution, our model has a compactness at a mass coordinate of $2.5 M_\odot$ of $\xi_{2.5} = 0.85$, placing it far above the threshold compactness for a failed explosion. 

However, the compactness of a star is an incomplete and debated explodability predictor. For example, recently, \citet{boccioli2025} found that highly compact cores can actually result in successful shock revival due to the effects of neutrino heating. In addition to this one-parameter estimate, our model fails to meet the explosion criteria of \citet{muller2016}, which predicts properties of neutrino-driven explosions based on a semi-analytic model for stellar structure. Meanwhile, the model is predicted to explode based on the \citet{ertl2016} test, which employs a two-parameter representation of stellar structure. We also evaluate our model according to the criterion of \citet{wang2022}, which is based on the maximum ram pressure derivative. We find that it lies just past their threshold for explodability, but the validity of this estimator for a star much more massive than those used to calibrate their threshold is unclear.

\begin{figure*}
\includegraphics[width=\linewidth]{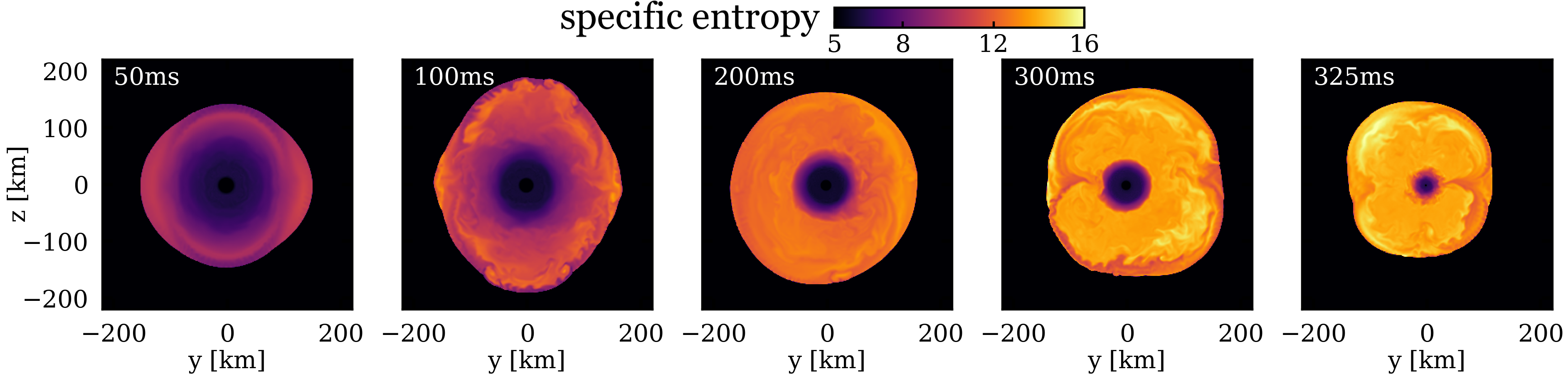}
\caption{Snapshots of the specific entropy $s$, in units of $k_B~\rm{baryon}^{-1}$, shown as 2D slices through our 3D \texttt{GRaM-X} simulation at various times (from left to right: 50, 100, 200, 300, and 325 ms post-bounce). The entropy generally increases from low values (blue) to higher ones (red) as time progresses. The contraction of the core region and the early shock expansion, followed by a later stalling, are also visible. The shock propagation and convective motions exhibit inherent asymmetries. A movie of the simulation is available at this link: \url{https://www.youtube.com/shorts/4sisdNGLUL0}.}
\label{fig:2d}
\end{figure*}

\subsection{GR1D Simulation} \label{sec:GR1D}

We map the pre-collapse model onto \texttt{GR1D} \citep{oconnor10}, a spherically-symmetric, general-relativistic neutrino hydrodynamics code, and evolve through the end of the core collapse process. \texttt{GR1D} is an open-source code\footnote{\url{https://github.com/evanoconnor/GR1D}} designed to simulate core-collapse and BH formation. It evolves the discretized general-relativistic hydrodynamics (GRHD) equations with a finite-volume scheme, piecewise-parabolic reconstruction, and a Riemann solver. Users can choose a microphysical, tabulated EoS and neutrino transport in the M1 formulation \citep{oconnor15} with tabulated multi-group neutrino opacities. \texttt{GR1D} also includes an approximate treatment of rotation, which renders the simulation effectively 1.5D. 

The initial conditions for the \texttt{GR1D} simulation are remapped from \texttt{MESA} at the onset of core-collapse, which we define as once the core infall velocity has reached $v_r > 1000~\mathrm{km}~\mathrm{s}^{-1}$. We choose a 1D grid that is uniform ($\Delta r = 100$ m) in the inner region (up to 2 km) and logarithmically spaced outside of it, with a total of 1200 radial zones (in addition to ghost zones). We map the enclosed mass, temperature, density, radial velocity, electron fraction, and angular velocity as functions of radial coordinate from \texttt{MESA} onto the new grid. The edge of the star on our grid is defined as where the density has dropped below $\rho_\mathrm{min} = 2000~\mathrm{g~cm}^{-3}$.

We choose a commonly-adopted EoS appropriate for hot nuclear matter from \citet{1991NuPhA.535..331L} with an incompressibility of $K_\mathrm{sat} =
220$ MeV (known as LS220). While LS220 is inconsistent with nuclear experimental data \citep[e.g.,][]{2017ApJ...848..105T}, we chose it for this first study for easier comparison with our and other previous simulation results. Our neutrino transport includes three species and 18 energy groups with tabulated opacities generated through the open-source neutrino interaction library NuLib\footnote{\url{http://www.nulib.org/}} that cover the vast parameter space of thermodynamic quantities.

Even with the effective rotation, spherically symmetric models will never adequately capture the details of the inherently multi-dimensional, asymmetric core-collapse process. However, key properties of the evolution agree well between the \texttt{GR1D} and our 3D GRMHD simulation with \texttt{GRaM-X}.

\subsection{GRaM-X Simulation} \label{sec:GRaM-X}
We map the same $M_\mathrm{ZAMS} = 45 M_\sun$ pre-collapse model onto the 3D grid of \texttt{GRaM-X}, our newly-developed GPU-accelerated dynamical-spacetime ideal-GRMHD code~\citep{Shankar2023}. \texttt{GRaM-X} has been developed for the \texttt{EinsteinToolkit} framework~\citep{Loffler_2012, roland_haas_2024_zenodo} and relies on the new adaptive mesh-refinement (AMR) driver \texttt{CarpetX}~\citep{schnetter_2022_zenodo, Shankar2023, Kalinani_2025}. We impose a moderately differential rotation profile and a weak magnetic field on top of the pre-collapse model. The angular velocity profile is given by 
\begin{equation}
    \Omega(r, z) = \Omega_0 \frac{r_0^2}{r^2 + r_0^2} \frac{z_0^4}{z^4 + z_0^4},
\end{equation}
where $r$ is the distance from the rotation axis in the equatorial plane, and $z$ is the distance from the equatorial plane. We choose parameters $r_0 = 500\, \mathrm{km}$, $z_0 = 1000\, \mathrm{km}$, and a core rotation velocity of $\Omega_0 = 0.44\, \mathrm{rad/s}$, which matches that of the evolved MESA model. We set up the initial pre-collapse magnetic field using a vector potential of the form
\begin{equation}
    A_r = 0, \\
    A_{\theta} = 0,\\
    A_{\phi} = B_0  \frac{r_0^3}{r^3+r_0^3} r \mathrm{sin} \theta
\end{equation}
where $A_r$, $A_{\theta}$ and $A_{\phi}$ are the $r$, $\theta$ and $\phi$ components of the magnetic vector potential. We choose parameters $r_0 = 2000\, \mathrm{km}$ and $B_0 = 10^{10}\, \mathrm{G}$.

We adopt the same grid setup as~\cite{shankar_2025}, so we describe it only briefly here. The initial simulation grid includes 4 AMR levels, each of which doubles the resolution. We progressively add 5 more AMR levels between the onset of core-collapse and PNS formation according to a set of density criteria. Our finest resolution is $370\, \mathrm{m}$. Once the shock forms, it remains within AMR level 6, with $1.48\, \mathrm{km}$ resolution. 

We use a 3D Newton-Raphson \citep[\textit{3DNR}, ][]{cerda_duran:2008} as the primary conservative-to-primitive transformation method and the method of Newman \& Hamlin \citep[\textit{Newman}, ][]{Newman_Hamlin:2014} as a fallback if primary method fails. In the post-bounce phase, we find that if the shock remains stalled for a few ms, undesirable high-frequency oscillations can develop in the magnetic field just outside the shock, leading to magnetohydrodynamic instabilities. To avoid this issue, we add diffusivity to the magnetic field \textit{outside} the shock using a modified Ohm's Law for the electric field calculation in our constrained transport implementation, given by~\citep{Moesta2015}
\begin{equation}
    \mathbf{E} = -\mathbf{v} \times \mathbf{B} + \eta \mathbf{J}, \\
    \mathbf{J} = \nabla \times \mathbf{B}
\end{equation}
where $\mathbf{E}$ is the electric field, $\mathbf{B}$ is the magnetic field, $\mathbf{v}$ is the 3-velocity and $\mathbf{J}$ is the 3-current density. We set $\eta = 0$ inside the shock and $\eta \sim 0.1$ outside the shock.

While our \texttt{GRaM-X} simulation uses a simple neutrino treatment, a leakage scheme with M0, as opposed to a more accurate scheme like the M1 transport implemented in \texttt{GR1D}, we do not expect a significant qualitative difference in the evolution. The progenitor core is relatively slowly rotating, rendering the multi-dimensional effects minimal, and our 3D evolution is consistent with the 1D simulation that includes sophisticated neutrino transport. We will be able to verify the detailed impact of this simplification when M1 is implemented in \texttt{GRaM-X}.

We also note that while evolving the MHD equations demonstrates the capability of our code for future applications, the weak field on the pre-collapse core in this case has negligible effects on the evolution. To verify this, we calculated plasma $\beta$, the ratio of gas pressure to magnetic pressure, which serves as a measure of the dynamical importance of the magnetic field. We find that $\beta > 100$ everywhere and $\beta \gg 100$ in most regions throughout the evolution.

\subsubsection{Apparent Horizon Finder} \label{sec:AHfinder}
To estimate the properties of the BH immediately after formation, we use an apparent horizon finding algorithm based on \texttt{AHFinderDirect} \citep{AHFinderDirect} as implemented in the \texttt{EinsteinToolkit} framework~\citep{Loffler_2012, roland_haas_2024_zenodo}. The apparent horizon corresponds to the ``marginally outer trapped surface'' (MOTS), the outermost surface through which outgoing null geodesics have no expansion. The apparent horizon serves as a dynamical probe of BH position and shape in a snapshot of an actively evolving simulation, from which the event horizon cannot be determined. It is contained within the event horizon and equals it exactly for a stationary spacetime. 
% Mathematically, the apparent horizon is described by an elliptic partial differential equation, which \texttt{AHFinderDirect} solves directly using fourth-order finite differencing and Newton's method.

\begin{figure}
\includegraphics[width=\linewidth]{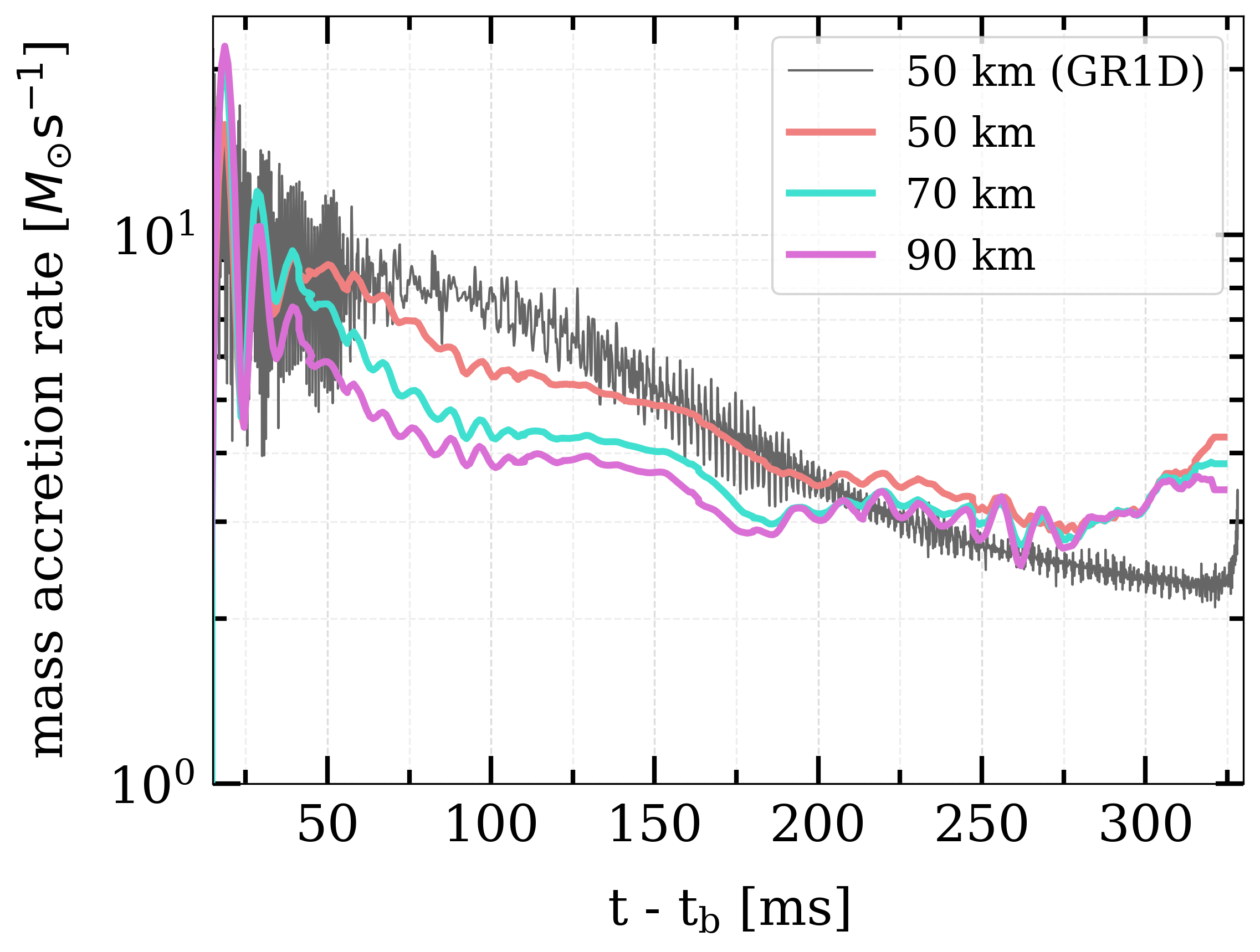}
\caption{The evolution of mass accretion rate, measured in solar masses per second, after core bounce. For the 3D \texttt{GRaM-X} simulation, we show the accretion rate measured through 3 different radii: 50 km (coral), 70 km (turquoise), and 90 km (pink). At early times, the accretion rate is higher through smaller radii, but at late times, the rate converges at a few solar masses per second. We also include the accretion rate measured at 50 km from our \texttt{GR1D} simulation (gray) for comparison; it shows a similar trend to the 3D results but ends with overall lower accretion.}
\label{fig:Macc}
\end{figure}

\section{Results} \label{sec:results}

Our 3D GRMHD simulation reaches core bounce at $t_b=341.2~\rm{ms}$, after which a shock forms, stalls, and falls back. Figure \ref{fig:2d} shows 2D slices of specific entropy at several times during the subsequent evolution. In the initial post-bounce phase (left panel of Figure \ref{fig:2d}), sloshing motions, likely a signature of the Standing Accretion Shock Instability \citep[SASI][]{blondin:03,laming:07,nagakura08}, are primarily along the Cartesian axes. The initially expanding shock stalls at a radius of $\approx~160~\rm{km}$ after about 100 ms (second panel of Figure \ref{fig:2d}). The shock evolution, including its oscillatory behavior and angle-dependence, is also visible in the central panel of Figure \ref{fig:1dcomparison}, which includes the shock radius averaged both in the polar and equatorial planes.

Over time, sloshing motions continue with the average shock radius remaining roughly static, while entropy increases significantly. After about 200 ms, a spiral mode begins to dominate the sloshing motions (central panel of Figure \ref{fig:2d}). This spiral mode is also a probable manifestation of the SASI \citep{Fernandez_2010}, which is known to play a key role in the dynamics and outcome of CCSNe \citep[e.g.,][]{yamasaki:06,ohnishi:06,marek:09,hanke:12,hanke:13,fernandez:15a}
Throughout this stage of evolution, the accretion rate, which began at a high value of $\approx 10~\mathrm{M_\odot~s}^{-1}$, drops to a value of roughly $3~\mathrm{M_\odot~s}^{-1}$, as seen in Figure \ref{fig:Macc}. At late times, approximately 250 ms post-bounce, the shock reverses (central panel of Figure \ref{fig:1dcomparison}) and the accretion rate rises once again (Figure \ref{fig:Macc}). The central density and PNS mass both rise steadily throughout the post-bounce evolution (top and bottom panels of Figure \ref{fig:1dcomparison}).

\begin{figure}[ht!]
\includegraphics[width=\linewidth]{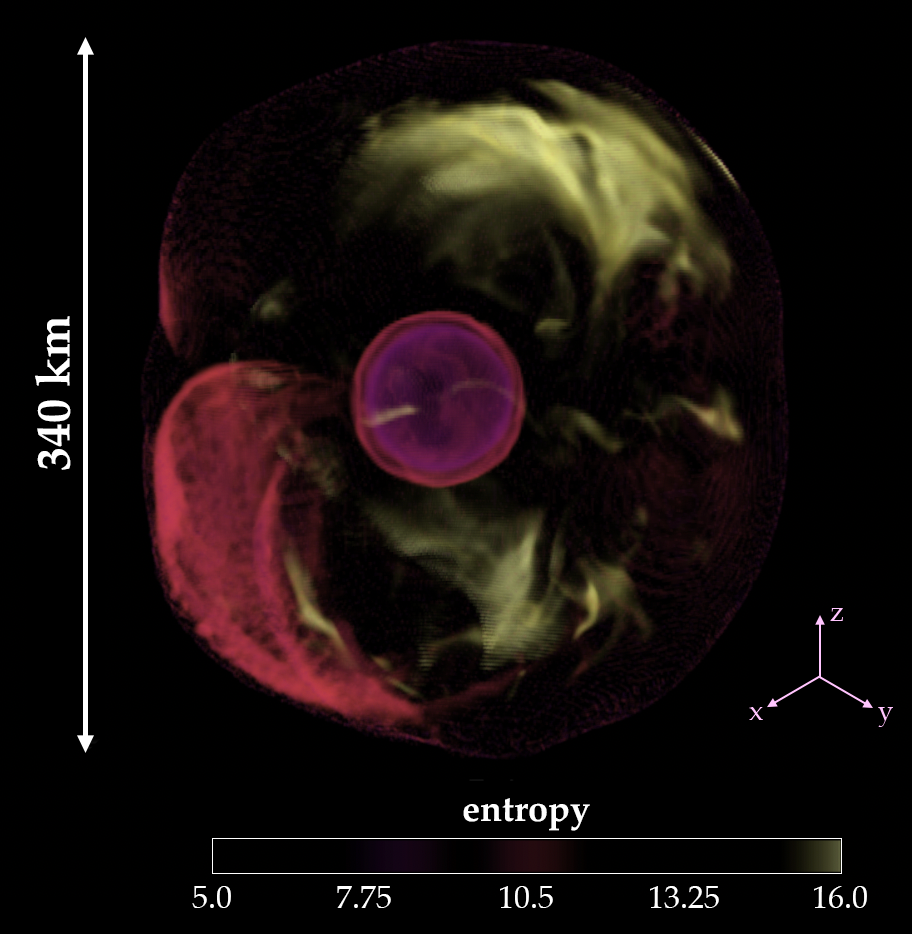}
\caption{Volume rendering of specific entropy $s$, in units of $k_B~\rm{baryon}^{-1}$, at 300 ms after core-bounce (corresponding to the 4th panel of Figure \ref{fig:2d}).}
\label{fig:3d}
\end{figure}

At late times, the PNS accretes lower entropy material, primarily in the equatorial plane (rightmost panels of Figure \ref{fig:2d}). In Figure \ref{fig:3d}, we show a volume rendering of specific entropy at 300 ms post-bounce, corresponding to the second to last panel of Figure \ref{fig:2d}. The asymmetry of the stellar material surrounding the PNS is evident in both the 2D and 3D snapshots.

The final panel of Figure \ref{fig:2d} shows the state of the simulation immediately before BH formation. At this time, the shock radius has decreased significantly (see the central panel of Figure \ref{fig:1dcomparison}, too) and the low-entropy PNS has also begun to collapse. At this point, the accretion rate has risen to $\approx 4~\mathrm{M_\odot~s}^{-1}$. The final collapse to BH formation is reflected by a dramatic, rapid rise in central density (top panel of Figure \ref{fig:1dcomparison}). Following this sharp rise, we identify the apparent horizon of the newborn BH and track its properties for a short time (see \S \ref{sec:BHprops}).

\subsection{Comparison to GR1D} \label{sec:comp_1d}
\begin{figure}[ht!]
\includegraphics[width=\linewidth]{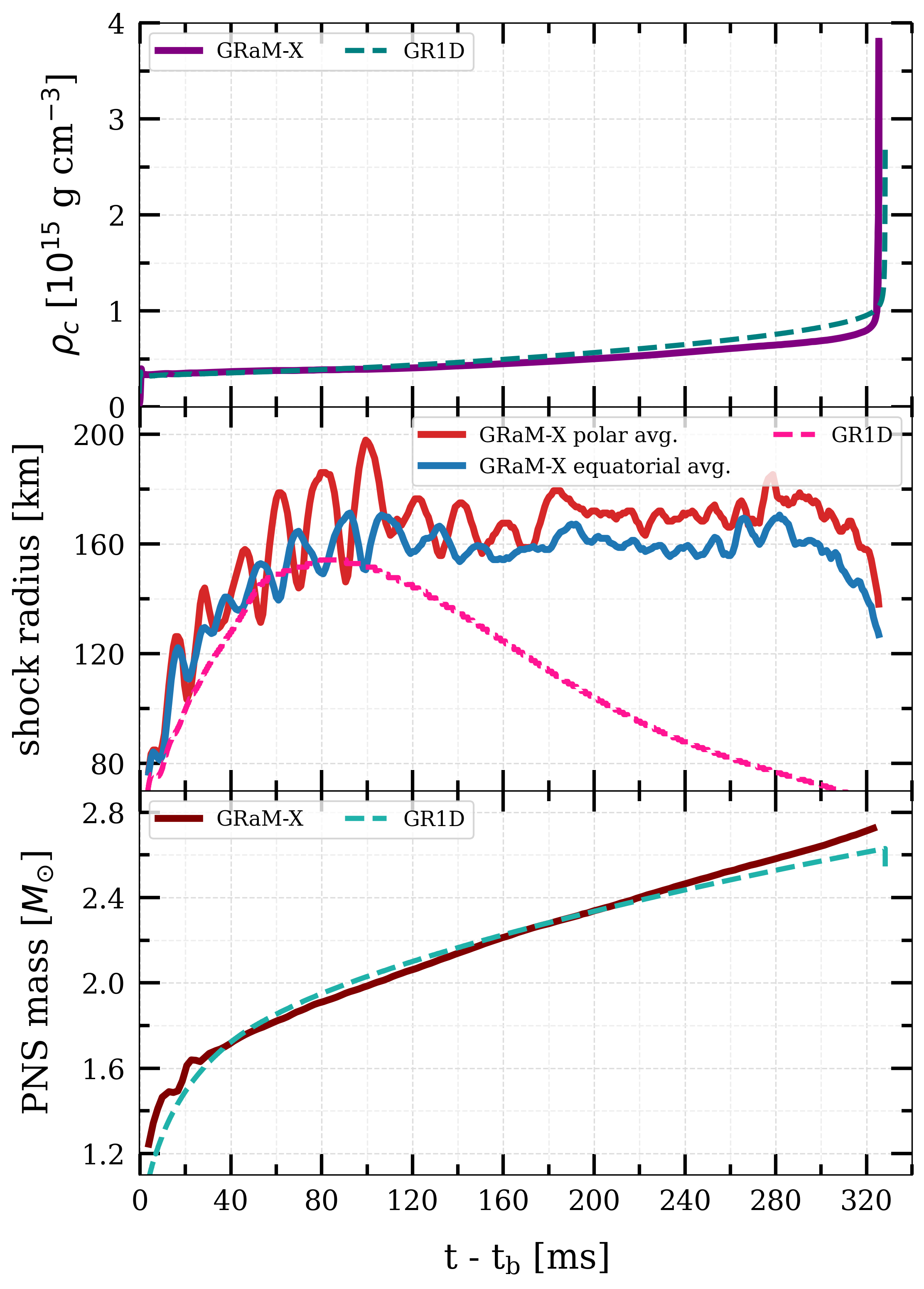}
\caption{Post-bounce evolution of the central density (top panel), shock radius (middle panel), and PNS mass (bottom panel) for both our \texttt{GR1D} simulation (dashed lines) and our 3D \texttt{GRaM-X} simulation (solid lines). The 3D evolution of the central density and PNS mass are both closely approximated by the 1.5D evolution. The time of BH formation, where the central density exponentially rises, is also nearly identical between the two simulations. However, the shock radius differs dramatically between the 3D and 1.5D evolution at times after it has reached its maximum value.}
\label{fig:1dcomparison}
\end{figure}

In \texttt{GR1D}, core-bounce occurs at $t_b=285.3~\rm{ms}$, significantly earlier than in 3D. This difference of about 44 ms is not particularly surprising given that \texttt{GR1D} uses an approximate treatment of rotation and a more sophisticated, multi-group treatment of neutrino transport. Given the differences between \texttt{GR1D} and \texttt{GRaM-X}, the similarity in their post-bounce evolution is remarkable. BH formation in \texttt{GR1D} occurs at $t-t_b =328.25~\rm{ms}$, compared to $t-t_b =325.42~\rm{ms}$ for the same model evolved in 3D GRMHD. Figure \ref{fig:1dcomparison} compares the evolution of central density, shock radius, and PNS mass between the 3D and 1.5D simulations. Overall, we find strong agreement between the two methods in terms of the central density (top panel) and PNS mass. The central density in \texttt{GR1D} rises slightly faster before BH formation. The PNS mass rises more smoothly in 1.5D, while it shows features due to episodic accretion in the \texttt{GRaM-X} simulation.

The shock radius, a fundamentally multi-dimensional quantity, shows similar, though far smoother evolution in 1.5D until roughly $80~\rm{ms}$ post-bounce. At this point in the \texttt{GR1D} simulation, the shock stalls and turns around immediately after reaching its maximum value of $r\approx 153~\rm{km}$. Meanwhile, in 3D, the shock radius oscillates significantly throughout the simulation, especially at early times, and remains at a similar radius from $80~\rm{ms}$ until its reversal at $\approx 280~\rm{ms}$. The nature of oscillations in 3D changes as the sloshing motions become dominated by spiral modes. 

\subsection{BH natal properties} \label{sec:BHprops}

Tracking the apparent horizon enables us to extract quasi-local BH properties, such as mass and kick, at early times, allowing for estimates of BH parameters without evolving them to their asymptotic values. We use a horizon-finding algorithm implemented in \texttt{GRaM-X}, as described in \S \ref{sec:AHfinder}, to identify the horizon and track its evolution for a brief time following BH formation. We evolve our simulation for a time $t_\mathrm{f}-t_\mathrm{BH}\approx 47.3~\mu\rm{s}$ following BH formation. Continuing to stably evolve the MHD solution requires developing a more careful treatment of the region within the apparent horizon, where gradients of the geometry steepen and can cause numerical issues.

\begin{figure}[ht!]
\includegraphics[width=\linewidth]{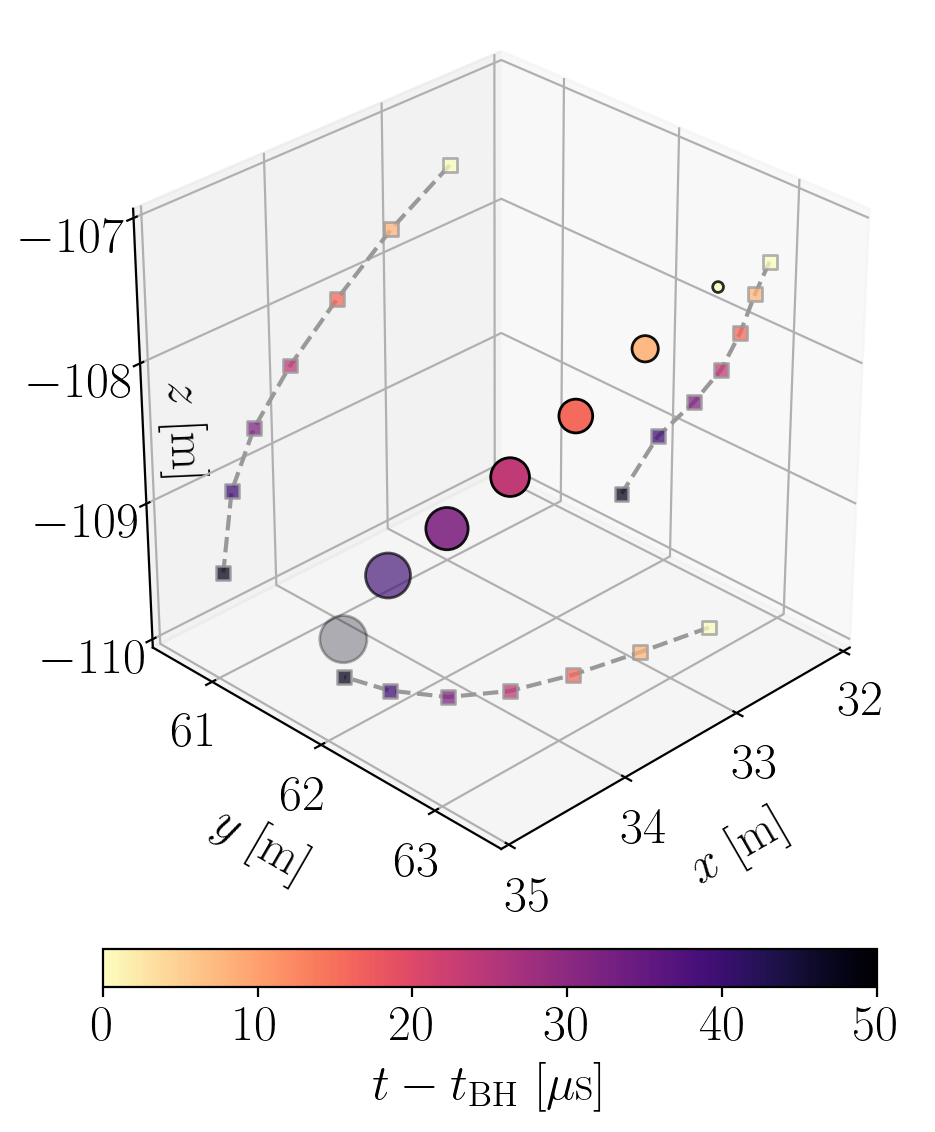}
\caption{Spatial trajectory of the BH as measured by the centroid of the apparent horizon over time. Circle markers in the 3D projection represent the horizon at each time after BH formation, with color indicating time after BH formation $t_\mathrm{BH}$ in microseconds and size representing the proper area $A$ of the horizon. The horizon expands at a decreasing rate while the black hole moves along a complex trajectory. 2D projections of this trajectory on each plane are drawn with square markers and connected with dashed gray lines.}
\label{fig:BHtrajectory}
\end{figure}

The most informative quantities we extract from \texttt{AHFinder} are the horizon centroid position, proper area $A$, and corresponding irreducible mass $M_\mathrm{irr} = \sqrt{A/16\pi}$. After its initial formation, the BH shifts due to kicks from the asymmetric accretion process. While neutrino asymmetries are also present, we expect them to be highly subdominant compared to the fluid effects. Figure \ref{fig:BHtrajectory} shows the 3D path of the BH up until the end of our simulation. We measure the kick velocity directly as
\begin{equation}
    \vec{v}_\mathrm{kick} = \frac{\delta x}{\delta t}\hat{i} + \frac{\delta y}{\delta t}\hat{j} + \frac{\delta z}{\delta t}\hat{k}.
\end{equation}
The kick velocity in each direction evolves with time. It decreases in the $+\hat{i}$ direction while increasing in the $-\hat{j}$ and $-\hat{k}$ directions overall. The mean velocities in the $\hat{i}$, $\hat{j}$, and $\hat{k}$ directions are $\approx 43$ , $-26$, and $-45~\mathrm{km/s}$, respectively. The total magnitude of the kick velocity vector has a mean value of $\left<v_\mathrm{kick}\right> \approx 72~\mathrm{km/s}$ and remains nearly constant during the short post-BH formation evolution. We note that the kick at this early time may still change significantly due to subsequent evolution \citep[e.g., via tug-boat mechanism,][]{janka2017}.

The size of each marker in Figure \ref{fig:BHtrajectory} corresponds to the area $A$, and can be seen to increase with time, plateauing by the end of the simulation. Since the irreducible mass $M_\mathrm{irr}$ scales like the square root of $A$, this corresponds to an increasing mass as well. The mass rises from its initial value of $M_\mathrm{irr}(t_\mathrm{BH}) \approx 2.55M_\odot$ to a final value of $M_\mathrm{irr}(t_\mathrm{f}) \approx 2.62M_\odot$ at $t_\mathrm{f}-t_\mathrm{BH}\approx 47.3~\mu\rm{s}$. We note that this mass is not identical to the local BH mass, but serves as a useful proxy. To extrapolate $M_\mathrm{irr}$ to later times, we fit a logarithmic function of the form,
\begin{equation}
    M_\mathrm{irr}(\tau) = C \log_{10}(\tau+\tau_0)+M_\mathrm{0}
\end{equation}
where $\tau \equiv t-t_\mathrm{BH}$ in units of $\mu \rm{s}$, to the early-time data. The best-fit values for $C$, $\tau_0$, and $M_0$ are 0.12472, 17.6472, and 2.3941, respectively. Hence, the mass is increasing logarithmically at the end of our simulation and is expected to continue to rise with time. If we extrapolate this fit forward in time, the mass reaches $M_\mathrm{irr}\approx 3.14M_\odot$ one second after BH formation.

Calculating the angular momentum in the PNS immediately before collapse enables an indirect estimate of the newborn BH's spin. We find that the total angular momentum in the PNS, defined as where the density exceeds $\rho_\mathrm{PNS} = 10^{11}~\mathrm{g~cm}^{-3}$, is only $\sim 5 \times 10^{47}~\mathrm{erg~s}$. This corresponds to a specific angular momentum of $\sim 9\times 10^{13}~\mathrm{cm}^2~\mathrm{s}^{-1}$ and a dimensionless spin that is less than 1\% of the maximal value for our BH mass (i.e., dimensionless spin $a < 0.01$). However, we note that accretion in the seconds after BH birth may lead to substantial spin-up.

\section{Discussion and Conclusion} \label{sec:discussion}
In this Letter, we present a proof-of-concept result: we can, for the first time, self-consistently simulate BH formation as the product of the stellar core-collapse process with a full 3D GRMHD simulation. To achieve this, we employ \texttt{GRaM-X}, a novel code that harnesses the benefits of GPU acceleration and abundant GPU resources, rendering a long-duration, resolved 3D dynamical-spacetime GRMHD simulation tractable. This capability is crucial for our understanding of the endpoints of stellar evolution on the high-mass end of the initial mass function. The total cost of $\sim3600$ node-hours is less than 1\% of a typical INCITE/tier-0 allocation for a given year, demonstrating the potential for more GPU-accelerated long-term full-GR simulations capturing BH formation. The slowdown associated with the planned implementation of more accurate neutrino transport (M1) is expected to be only a factor of a few, leaving simulations with the neutrino treatment necessary for nucleosynthetic calculations accessible.

In future work, we will continue to evolve the simulation after BH formation for a longer duration to get a better grasp on the final BH mass and kick. This requires carefully treating the region within the apparent horizon to maintain stability in the MHD solution, but it is within reach as an extension of the early work presented here. We will also extract more information from the apparent horizon to probe BH spin, another key property to connect to GW observations. In this Letter, we applied our GRMHD code, \texttt{GRaM-X}, to the case of a weakly magnetized model, where the field is not dynamically important. For this first study, we chose these initial conditions to optimize for rapid BH formation. However, we can also capture BH formation from a progenitor star with a strong core magnetic field and rapid rotation, which has particular relevance for understanding the connection between lGRB and type Ic-bl SNe. This capability is beyond that of any other code due to either missing physics (e.g., lack of full GR, MHD, or 3D effects), excessive computational expense, as taking advantage of GPU acceleration makes \texttt{GRaM-X} unique among numerical relativity MHD codes, or both. Finally, we can remap the end state of a simulation as presented here to a stationary metric and follow the accretion disk and jet launching from self-consistent initial conditions, to answer questions such as whether collapsars can drive type Ic-bl supernovae, lGRBs, and contribute to the nucleosynthesis of \textit{r}-process elements \citep{2025arXiv250202077S, issa2025}.

Moving forward, we are no longer limited to sacrificing multidimensionality or forced to extrapolate properties of the compact object remnant from shorter 3D simulations. This result opens the door to more such simulations, including parameter studies, which can shed light on the natal distributions of stellar-mass BH masses, spins, and kicks. Such information has synergies with the relatively new field of gravitational wave observations of BH mergers. Our results also lend confidence to those obtained from 1D parameter studies performed with \texttt{GR1D}. For example, \citet{2020ApJ...894....4D} explores the formation of stellar-mass BHs in spherically symmetric simulations while varying the EoS and progenitor stars. The accuracy of the BH properties predicted from \texttt{GR1D}, as validated by our 3D GRMHD simulation, suggests that trends gleaned from such studies hold weight.

We believe there is progress to be made in a complementary approach that combines wider parameter studies in reduced dimensionality with full 3D dynamical-spacetime simulations of a few special cases, like the one we present here. The opportunity to directly measure natal BH properties renders such simulations uniquely valuable to furthering our understanding of the relationship between stellar core-collapse and compact objects.

\begin{acknowledgments}
GH recognizes support from a Visiting Scholarship at CIERA/Northwestern University. This research was supported in part by grant NSF PHY-2309135 to the Kavli Institute for Theoretical Physics (KITP), and in particular, the 2024 ``Turbulence in the Universe'' program attended by GH and PM. We would like to thank the anonymous reviewer for comments which improved this manuscript.
SS would like to thank Sherwood Richers for hosting him as a post-doctoral fellow at the University of Tennessee in Knoxville where some of the work was carried out. 
This work has benefited from participation in the \texttt{GRaM-X} Hackathon 2024 at the Perimeter Institute for Theoretical Physics. 
PM acknowledges funding through NWO under grant No. OCENW.XL21.XL21.038.
RH acknowledges support by NSF awards OAC-2004879, OAC-2005572, OAC-2103680, OAC-2310548, OAC-2411068. 
ES acknowledges the support of the Natural Sciences and Engineering Research Council of Canada (NSERC). Research at Perimeter Institute is supported in part by the Government of Canada through the Department of Innovation, Science and Economic Development and by the Province of Ontario through the Ministry of Colleges and Universities. 
This research used resources of the Oak Ridge Leadership Computing Facility at the Oak Ridge National Laboratory, which is supported by the Office of Science of the U.S. Department of Energy under Contract No. DE-AC05-00OR22725. The simulations were carried out on OLCF’s Frontier using the INCITE-2024 award under allocation
AST191.

\end{acknowledgments}

\vspace{5mm}

\bibliography{merged}{}
\bibliographystyle{aasjournalv7}

\end{document}